\documentstyle[12pt]{article}

\def\tag#1{\space}%
\def\text#1{\space}%

\begin{document}

\begin{flushright}
 {\large DAPNIA/SPP 95-15} \\
 {\large SPhT-T95-090 } \\
\end{flushright}

\vspace{1cm}
\begin{center}
  \begin{Large}
  \begin{bf}
Deep-Inelastic Onium Scattering \\
  \end{bf}
  \end{Large}
  \vspace{5mm}
  \begin{large}
    $H. Navelet^{1}, R. Peschanski^{1}, Ch. Royon^{2}$ \\
  \end{large}
\vspace{1cm}
1- Service de Physique Th\'eorique, Centre d'Etudes de Saclay \\
F-91191 Gif-sur-Yvette Cedex (France)\\
2- DAPNIA - SPP, Centre d'Etudes de Saclay\\
F-91191 Gif-sur-Yvette Cedex (France)\\
  \vspace{5mm}
\end{center}
\begin{quotation}
\noindent
{\bf Abstract:}
Using the colour dipole approach of the QCD perturbative (BFKL) Pomeron
exchange in onium-onium scattering, we compute the cross section for
small
but hierarchically different
onium sizes. A specific
 term  dependent on the size-ratio is generated. In deep inelastic
 onium scattering  it appears as
  a scaling-violation contribution to the quark
structure function
  near the BFKL singularity. We find that the
extension of the formalism for deep inelastic onium  scattering
 to the proton structure function provides a
remarkably good 3-parameter fit to HERA data at  small $x$
with a simple physical interpretation in terms of the dipole
formulation.

\end{quotation}

\newpage\ Recently, a dipole model approach to the QCD computation
 of onium-onium scattering from the  perturbative resummation of leading
 logarithms
 has been proposed$^{\cite{1,2,3}}.$
 In this framework, each onium state describes a small (i.e.
massive) colour-singlet quark-antiquark pair whose infinite momentum
wavefuntion
 is
itself composed of colour dipoles at the moment of the interaction.\ These onia
 interact locally in impact-parameter space
via 2-gluon exchange between the component dipoles.
 The overall total cross-section has been shown to reproduce the
well-known$^{\cite{4}}$ QCD Pomeron exchange (BFKL) due to leading $\ \log
1/x$ resummation. It is thus tempting to try and apply this formalism to
 the deep-inelastic scattering off protons and more specifically to the
proton  structure function measured at HERA$^{\cite{5}}$.

However, before  applying the dipole formulation to the experimental
case, a few  problems should be treated. First, it is important to take into
account the disymmetry between the scales involved in deep-inelastic
scattering,
namely the photon virtuality $1/Q$ and the scale  typical of the target.
It is thus mandatory to treat the dipole-dipole cross-section for
different initial scales $X_{01} \not = X_{02},$ while maintaining
both scales small, i.e. in the perturbative regime. The resulting formulae
are derived in section ${\bf 1}.$

Second, we have to  convert to the
case of deep-inelastic scattering of a virtual photon on an onium state, itself
considered as a certain distribution of initial dipole configurations.  In
order
to describe correctly the coupling of the photon to quarks in the dipole
 formulation$^{\cite{1}},$
it is necessary to invoke the $k_T -$factorization scheme$^{\cite{6}}.$
This amounts  introducing a modification in the dipole-dipole cross-section
formula, with the substitution  $X_{10} \to 1/Q$ and $X_{20} \to 1/Q_0,$
where $Q_0^{-1}$ is of the order of the average dipole size in the
 onium
state. This derivation is performed in section {\bf 2.}

Finally, one would like to translate the previous results obtained for
 deep-inelastic onium
scattering  into the experimentally accessible deep-inelastic proton
 scattering.
As already well-known  from the discussion of the BFKL contribution, this step
requires  assuming a negligeable contamination of
non-perturbative contributions to  the BFKL
 singularity. We shall
make the same kind of assumption by considering the proton at small $x$ as
composed of dipoles interacting with the photon in the same way as the for the
onium case. Within this approximation, we can apply our formulae to proton
deep-inelastic scattering at small $x$ in the kinematical
range at which it is performed at HERA.
We make the  comparison with data in section {\bf 3.}
We first determine the best fit of our formula for the published set of data
on the proton structure function at $x < 10^{-2}$, with only 3 free parameters
 for
100 experimental points, and then we confront our predictions with the
most recent set of  data at lower $Q^2.$

{\bf 1.} Starting with incident onia whith a-priori different sizes $X_{01}$
and $X_{02}$ (corresponding to masses of order ${X_{01}^{-1}},{X_{02}^{-1}}$),
one writes$^{\cite{2}}$
 for the total scattering
cross-section at fixed impact-parameter $b:$

\begin{eqnarray}
\frac{d^2\sigma }{d^2b} &=&\int \int \frac{d{X_1}}{X_1}\frac{%
d{X_2}}{X_2}\ \sigma \left( X_1,X_2\right) \label{un} \\
&\times& \int d^2b_1\ d^2b_2\ d^2%
b\ \delta ^2\left( b-b_1-%
b_2\right)\ \nonumber \\
&\times &\ \eta \left( X_{01};b_1,X_1;Y/_2\right)
\ \eta \left( X_{02};b_2,X_2;Y/_2\right) ,  \nonumber
\end{eqnarray}

\noindent where

\begin{equation}
\sigma \left( X_1,X_2\right) =4\pi \alpha ^2\ \int_0^\infty \frac{d\ell
}{\ell ^3}\left[ 1-J_0\left( \ell {X_1}\right) \right] \left[ 1-J_0\left(
\ell {X_2}\right) \right]   \label{deux}
\end{equation}

\noindent is the 2-gluon exchange amplitude between dipoles of size
respectiveley  $X_1$%
{\rm \ and }$X_2,$ and

\begin{eqnarray}
\eta \left( X_{0i};b_i,X_i;Y/_2\right) &\simeq &\frac{X_{0i}}{X_i%
}\frac 1{\pi b_i^2}\int_{-\infty }^{+\infty }\frac{d\nu _i}\pi \times
\nonumber
\label{trois} \\
&\times & 2i\nu _i\ \left\{ \frac{ b_i^2}{X_{0i}X_i}%
\right\} ^{2i\nu _i}\exp \left\{ \frac{N_c\alpha _s}\pi \chi \left( \nu
_i\right)  Y/ _2\right\} ,  \label{trois}
\end{eqnarray}

\noindent is the multiplicity distribution of dipoles for given
 initial
 size $X_{0i},$ impact parameter $b_i,$ size $X_i$ and
rapidity $Y/_2.$

As usual,
\begin{equation}
\chi \left( \nu _i\right) =2\Psi (1)-\Psi \left( \frac
12+2i\nu _i\right) -\Psi \left( \frac 12-2i\nu _i\right)\  ;\ \Psi
{(\gamma )}\equiv \frac{d\log \Gamma }{d\gamma },
\end{equation}
\noindent denotes the Mellin-transformed BFKL kernel.
 Note that formula (\ref{trois}) can be obtained
from the wave functionale
at infinite-momentum
$^{%
\cite{2}}$
 within the approximations $
b_i/X_i, b_i/X_{0i}\ll 1,\nu _i\approx 0.$

Inserting expressions (2) and (3) inside formula (1), let us proceed
further with a
determination of the cross-section keeping the two scales $X_{01}${\rm \ and
}$X_{02}$ small, but assuming a rather large ratio
${X_{01} \over
X_{02}}.$
 We first perform the independent integrations on $b_1,b_2${\rm \
and }$\ell $ to get:

\begin{eqnarray}
\frac{d^2\sigma }{d^2b} &\simeq &4\ \frac{\alpha ^2}{b^2}%
X_{01}X_{02}\int \int \frac{d{X_1}}{X_1^2}\frac{d{X_2}}{X_2^2}\ \frac
{X_{<}^2}4\left[ 1+\ln\left( \frac{X_>}{X_<}\right) \right]
\label{quatre} \\
&\times & \int \int \frac{d\nu _1}\pi \frac{d\nu _2}\pi \left( 2i\nu _1+2i\nu
_2\right) \left( \frac{b^2}{X_{01}X_1}\right) ^{2i\nu _1}\left( \frac{b^2}{%
X_{02}X_2}\right) ^{2i\nu _2}\nonumber \\
&\times &\exp\left({\frac{N\alpha Y}{2\pi} \left( \chi (\nu
_1)+\chi (\nu _2)\right)} \right),  \nonumber
\end{eqnarray}

\noindent where $X_<\
{\rm {(resp.X}}_>{\rm )}$ is the smaller (resp. larger) of $
X_1$ and $ X_2.$ Note that both values are required to verify $X  < b.$
 After further integration over $X_{1},X_{2}$ (exact) and over $\nu
_1,\nu _2$ (saddle-point) one gets:

\begin{eqnarray}
\frac{d^2\sigma }{d^2b} &=&4\ \alpha ^2\ \frac{X_{01}X_{02}}{%
\pi b^2}\;a \ \exp \left( \left( \alpha _p-1\right) Y-a \ \left( \ln ^2\frac
b{X_{01}}+\ell n^2\frac b{X_{02}}\right) \right)  \label{cinq} \\
&=&4\ \alpha ^2\ \frac{X_{01}X_{02}}{\pi b^2}\;a \ \exp \left( \left(
\alpha _p-1\right) Y-\frac{a}2\ln ^2\frac{b^2}{X_{01}X_{02}}\right)
\nonumber \\ &\times& \exp\left({-%
\frac{a}2\ln ^2\frac{X_{01}}{X_{02}}}\right)  \nonumber
\end{eqnarray}

\noindent where, using the conventional notations, $\alpha _p$ is the
``intercept'' of the BFKL singularity and $a$ the ``diffusion''
coefficient at rapidity $Y^{\cite{4}}.$ One has

\begin{eqnarray}
\alpha _p-1 &=&\frac{\alpha N_c}\pi \chi _{(0)}\ \equiv \ \frac{\alpha N_c}\pi
4\ln 2  \label{six} \\
a \equiv a(Y) &=&\left[ -\frac{\alpha N_c}{4\pi }\chi ^{\prime \prime
 }(0)Y\right]
^{-1}=\left[ \frac{\alpha N_c}\pi 7\zeta (3)Y\right] ^{-1}.  \nonumber
\end{eqnarray}

\noindent The integrated cross-section $\sigma =\int d^2b\;\frac{%
d^2\sigma }{d^2b}$ reads:

\begin{equation}
\sigma =2\pi X_{01}X_2\ \alpha ^2\;e^{(\alpha _{p-1})Y}\left[ \frac{2a}\pi
\right] ^{1/2}\ \exp\left({-\frac{a}2\ln ^2\frac{X_{01}}{X_{02}}}
\right)  \label{sept}
\end{equation}

Our results summarized in formulae (\ref{cinq}) and (\ref{sept}) deserve some
comments. The essential feature
is the scale-ratio dependent factor

\begin{equation}
\exp \left({-\frac{a}2\ln ^2\left( \frac{X_{01}}{X_{02}}\right) }
\right)\equiv
\exp \left({-\frac
{\ln ^2\left( \frac{X_{01}}{X_{02}}\right)}{2\ K \left(\alpha_p-1\right)Y}}
\right),  \label{huit}
\end{equation}

\noindent where we have used eq.(\ref{six}) to define the constant $K$ such
 that:

\[
 a^{-1} \equiv K \left(\alpha_p-1\right)Y ; ~ \ K \equiv -\frac{\chi_{(0)}^
{\prime
 \prime }}{4\chi_{(0)}}=\frac{7\zeta
 {(3)}}{4\ln 2}%
\ \ \left( \sim 3\right) .
\]

\noindent When the scale ratio is equal to 1, one recovers
for $\sigma$ the known result\footnote{%
Note that the
expression (\ref{cinq}) for $\frac{d^2\sigma }{d^2b}$ is
formally different from the analoguous formula (\ref{dix}) of Ref.\cite{2},
even when $\frac{X_{01}}{X_{02}}=1.$ However the dominant contribution in (%
\ref{dix}), ref.\cite{2} is valid for
$a^{-1}\ \propto \ \ln \left( \frac{b^2}{X_{01}X_{02}}\right) ,
$ which restores the equivalence between the two expressions.}.

Expression (\ref{huit}) gives a non-trivial logarithmic dependence on
the scale independent variables typical of
the reaction namely $\ln \frac{X_{01}}{X_{02}}$ and $Y
.$

\noindent The factor (\ref{huit}) does play a role if the typical sizes of
the dipoles, while being both small in order to preserve the perturbative
treatment of the process, are hierarchically different
$\frac{X_{01}}{X_{02}} \gg 1.$ The interpretation of (\ref{huit})
as due to a genuine scaling-violation factor present in the dipole
derivation of
the BFKL singularity for deep-inelastic onium scattering
will be made clear in the forthcoming discussion. We shall also make
more precise
the range of validity of the scale-dependent factor.

{\bf 2.}
In order to describe deep-inelastic scattering on an onium state,
and in particular to fulfill the requirement of $k_T$ factorization$^{\cite{6}}
,$ let us introduce the unintegrated structure function $F^{\prime}
= \frac {dF(x,Q^2)}{d\ln Q^2}$ and its formulation$^{\cite 1}$ in terms of
the dipole distribution function inside the onium state. One writes$^{\cite
1}:$
\begin{equation}
xF^{\prime}(x) =2\frac {\alpha N_c}{\pi }\int_{1/2-i\infty }^{1/2+i\infty
 }\frac{%
d\gamma }{2i\pi }\left(\frac Q {Q_0}\right) ^{2\gamma}\ v_{\gamma}\
 e^{\frac{\alpha N}\pi \chi
(\gamma )\ln 1/x},  \label{dix} \\
\end{equation}
where $Q_0$ correspond to some average over the dipole size distribution
of the onium target (at least when $\gamma$ stays in the vicinity of
the critical value $\gamma_c =1/2).$ $v_{\gamma}$ is defined by
\begin{equation}
v_{\gamma} =\int_{1}^{\infty} v(u)\ u^{-2\gamma-1} du, \label{onze} \\
\end{equation}
where the function $v(u)$ describes the factorized vertex as a non-perturbative
 input in general. We only know that
$v(u) \to 1$ when $u$ becomes large enough.

Inserting (11) into formula (10), one finds the following expression:
\begin{equation}
F^{\prime} =2\frac {\alpha N_c}{\pi }\int_{1}^{\infty} v(u)\frac{du}{u}
\int_{1/2-i\infty }^{1/2+i\infty }\frac{%
d\gamma }{2i\pi }\left(\frac Q {u Q_0}\right) ^{2\gamma}\exp\left({\frac{\alpha
 N}\pi \chi
(\gamma )\ln 1/x}\right).  \label{douze} \\
\end{equation}

Now, let us suppose that we are interested to determine the Mellin integral
in a region where the ratio $Q/Q_0$ is large but not too large in order to
keep the $\gamma-$integration near $\gamma_c$ (these conditions will be made
quantitative further on). Then the convolution (12) will be dominated
by large values of $u \approx Q/Q_0$, and thus one may consider that
$v(u) \approx 1.$ The solution of (12) becomes straightforward, and after
further integration in $\ln Q^2,$ one gets the final answer:

\begin{equation}
xF(x) =\frac {\alpha N_c}{\pi }
\int_{1/2-i\infty }^{1/2+i\infty }\frac{%
d\gamma }{2i\pi\gamma^2}\left(\frac Q {Q_0}\right) ^{2\gamma}\  e^{\frac{\alpha
 N}\pi \chi
(\gamma )\ln 1/x}  \label{treize} \\
\end{equation}
and, using a saddle-point method by expansion around $\gamma_c,$ one obtains

\begin{equation}
xF(x) =\ \frac{\alpha N_c}{\pi} e^{(\alpha _{p}-1)\ln1/x}\left[ \frac{2a}\pi
\right] ^{1/2}\ \frac {Q}{Q_o}\ e^{-\frac{a}2\ln ^2\frac{Q}{Q_0}}
 \label{quatorze}
\end{equation}
with the variables $a, \alpha_p $ defined in the same way as in (7),
identifying
$Y \equiv \ln1/x.$ We thus recover a formula similar to (8), but now
appropriately defined in terms of the scale ratio $Q/Q_0$
where $Q$ is the photon virtuality and $Q_0,$ a typical scale
related to the average dipole size in the target.

Interestingly enough, the conditions to obtain formula (14) from the
 saddle-point method are such that they fix the conditions of its
 applicability.
Noting that the saddle point value is $\gamma^* = 1/2 - a\ln Q/Q_0,$ the
 consistent approximation leading to (14) is given by:

\begin{equation}
a\ln{Q/Q_0} \simeq \frac{\ln{Q/Q_0}}{\ln 1/x} \ll 1 ,\quad  a\ln^2Q/Q_0
\simeq  \frac{\ln^2Q/Q_0}{\ln 1/x}= O(1).
\label{quinze}
\end{equation}

One thus realizes that the proposed parametrization should be valid for the
region of moderate $Q/Q_0$ when compared to the range in $1/x.$ This makes it
an
 interesting  parametrization for the HERA range, provided one
may extend the validity of (14) from an hypothetical onium initial
state to the proton. This is the subject of the next section. Note that
at higher $Q^2$ values, one expects the usual Double-Log-Approximation
(DLLA) to become valid, which amounts$^{\cite{7}}$ to consider in
formula (13) the pole in
$1/\gamma$ in the kernel $\chi (\gamma),$ (see (4)). In that case,
however, the input function $v(\gamma)$ is not a-priori known.

Let us focus the discussion on the signification of the scale-dependent
factor (9) with respect to the usual derivations of the BFKL contribution.
Indeed, an inverse Mellin-transform similar to Eqns.(10-13) appears in
the classical derivations of the BFKL singularity \cite{4,8}, as well
as in the dipole-model formulation\cite{1,2}. In some cases, like
the production of a forward jet in deep-inelastic scattering\cite{9},
a similar scale factor, which depends on the ratio of the photon
virtuality to the jet transverse momentun, has been taken into account.
However, in that case, the physical goal was to emphasize the typical
scale-independent BFKL contribution by choosing this scale-ratio as possible
of $\cal {O}$(1) in order to damp the possible DGLAP evolution.

In our case, on contrary, the fact that both the dipole-model result
of Eqn.(8) and the inverse-mellin transform leading to Eqn.(14) give
similar results leads us to the conclusion that the scale-dependence
we obtain is a quite general feature of the BFKL singularity and
should be taken into account as a genuine scaling violation
prediction of the whole theoretical scheme. It is also an incentive
to extend our results from the original deep-inelastic onium
reaction to the more practical case of proton inelastic scattering
(with the assumption of neglecting non-perturbative effects).

{\bf 3.}
In order to test the accuracy of the parametrisation obtained above, a
fit using the published data of the H1 and Zeus experiments \cite {5} was
 achieved.
The parametrisation used for the fit is the following ( see Eqn. (14)):
\begin{equation}
F_{2}=C a^{1/2}\ \exp(1/Ka)\ \exp \left( \ln \frac{Q}{Q_{0}} (1-\frac{a}{2} \ln
\frac{Q}{Q_{0}} ) \right)
\end{equation}
where:
\begin{equation}
a^{-1} = K(\alpha_{P} -1) \ln \frac{1}{x}
\end{equation}
The parameters used in the fit are $\alpha_{P}$, $Q_{0}$, and $C$. The
data used for the fit were the published 93 data from the H1 and Zeus
experiments, with $x \leq 0.014$ and $Q^{2} \leq 250.GeV^{2}$
\cite {5}, which corresponds
to 100 measured points. This choice is motivated by  the
theoretical requirements, in particular the validity range defined
in (15).

The results of the fit are given in figure
{\bf 1}. It can be noticed that the high $x$ and $Q^{2}$
points were not present in the fit, (as expected, the behaviour
in that kinematical region is not reproduced by the theoretical curve).
The values of the parameters are
the following : $\alpha_{P} = 1.243$, $Q_{0}=0.513$, $C=0.090$, for a
$\chi^{2}$ equal to 93.4. If we compare with the values expected
by the theory, they are in fairly good agreement:, since one would expects
the following range of values:
$\alpha_{P} = 1+\frac{\alpha N_{C}}{\pi} 4 ln 2 \approx 1.3,$ and
$C=\frac{\alpha N_{C} }{\pi} \sqrt{\frac{2}{\pi}}\approx 0.1,$ while
the value of $Q_0$ corresponds to an average radius of $1GeV^{-1}$
for the dipole size in the proton which seems reasonable.
These values are obtained for $N_{C}=3$ and $\alpha = 0.12$.

 If one would have performed separate  fits using only the 60 H1 points
( with $x \leq 0.013$ and $ 4.5 \leq Q^{2} \leq 120$), or only the 40
Zeus points ( with $x \leq 0.014$ and $ 8.5 \leq Q^{2} \leq 250$),
we get respectively : $\alpha_{P} = 1.214 (1.351)$, $Q_{0}=0.516
(0.522)$,
$K=0.114 (0.043)$, and a $\chi^{2}=43.2 (27.2)$. All in all, the fit of the
data
 is
remarkably good and the values of the parameters are rather close to those
expected from the theoretical framework.
As foreseen, the high $Q^{2}$ predictions are not so good as
the parametrisation is not supposed to be valid in this domain, where the
DGLAP equation is supposed to be more accurate.
\par
To check the validity of
this parametrisation, the values of the measured $F_{2}$ obtained by the
H1 collaboration with the 1994 data (\cite {5}) were compared with the
parametrisation. It must be noticed that the parameters used in the comparison
 are
kept the same as for the previous fit without new adjustment, and simply
 compared
 with the
new data at lower $Q^{2}$. The comparison is shown in the figures {\bf 2},
and the agreement between the measured points at low
$Q^{2}$ ($Q^{2} \geq 2 GeV^{2}$) is perfect.

In conclusion, applying the dipole model to deep-inelastic onium
scattering, we have found the following results:

1) The dipole-dipole scattering cross-section between dipoles of
unequal masses exhibit a non-trivial factor dependent of the ratio of the
dipole sizes,

2) A similar factor appears in the structure function describing
deep-inelastic scattering on an onium state. It depends on the
ratio $Q/Q_0$, where $Q$ is the virtuality of the photon
and $Q_0^{-1}$ is related to the average size of the dipole
configurations of the onium. It plays the r\^ ole of a genuine
scaling-violation contribution associated to the BFKL singularity

3) Extending the model to deep-inelastic proton scattering,
we find a remarkably good description of the recent HERA data
on the proton structure function. A 3-parameter fit gives a $\chi^2$
value of less than 1 per point for the published H1 and ZEUS data,
while the extrapolation of the resulting parametrization to the
very recent low$-Q^2$ data is excellent. The 3 parameters found in the
fit stick to the values expected from the theoretical framework

 In view of the striking agreement between the theoretical dipole
 picture and the phenomenological description of the data at small $x,$
 we are led to think that the correct understanding of the proton
 constituent picture in this region requires the dipoles as the fundamental
 objects present during (and responsible of) the interaction.

\noindent {\bf Acknowledgements}

\noindent We would like to thank A.\ Bialas for the
motivation leading to this work, and J. Bartels, G. Salam,
S. Wallon for
many stimulating and helpful discussions
before and during the Cambridge Workshop on Physics at HERA held at Christ
 College in July 1995.

\begin{center}

{\bf Figure Captions}

{\bf Fig.1}

3-parameter fit of the proton structure function at HERA.

The fit is constrained by the 93 experimental data from the
H1 and ZEUS experiments, /cite{5} . The kinematical range
selected for the fit is $x < 0.014 Q^2 < 250 GeV^2,$ while the
whole set of data is shown on the figures. The black (resp. white)
points are the H1 (resp. ZEUS) points.

{\bf Fig.2}

prediction for the 1994 H1 data

The comparison is made between the resulting parametrization
obtained from the fit on the published 1993 data with the new
1994 data obtained at lower $Q^2$ by the H1 Collaboration.
\end{center}

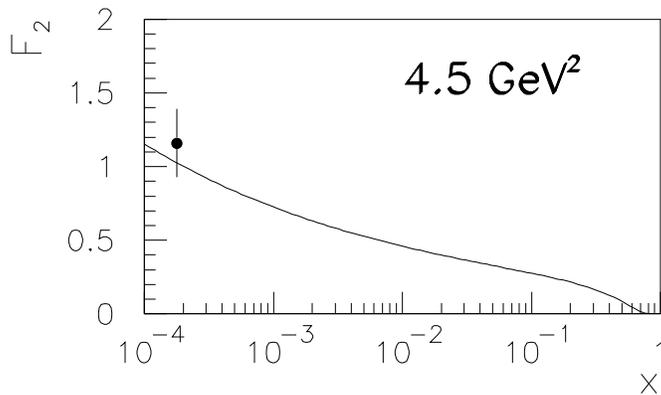
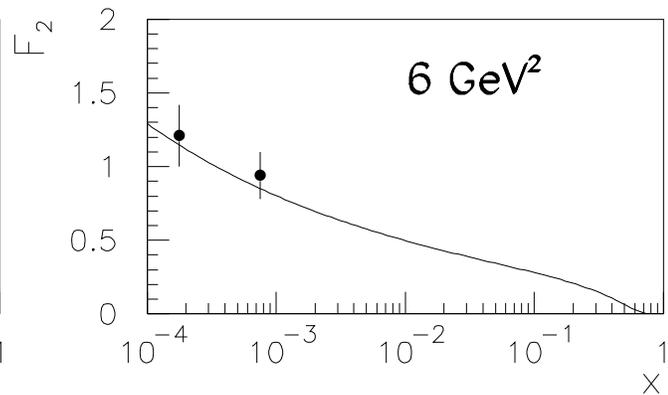
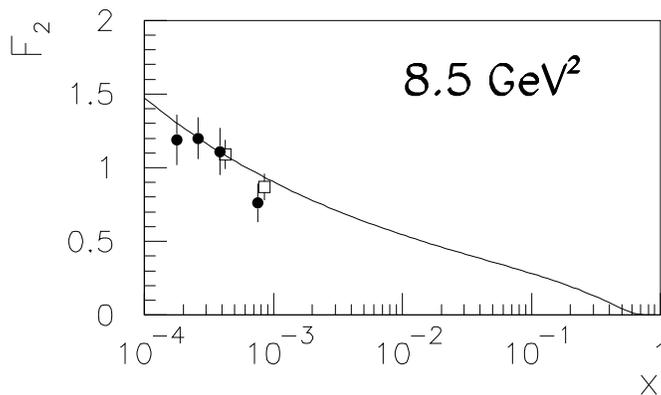
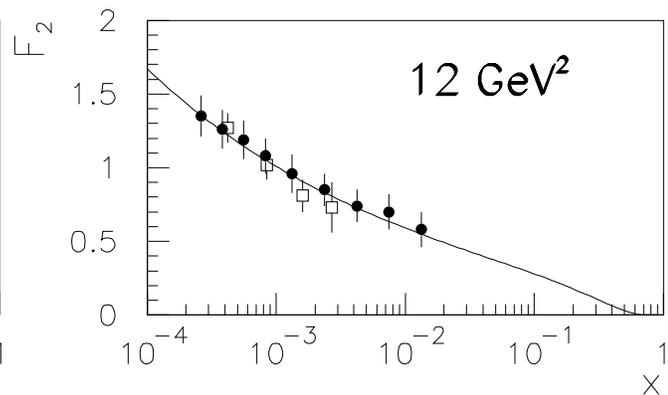
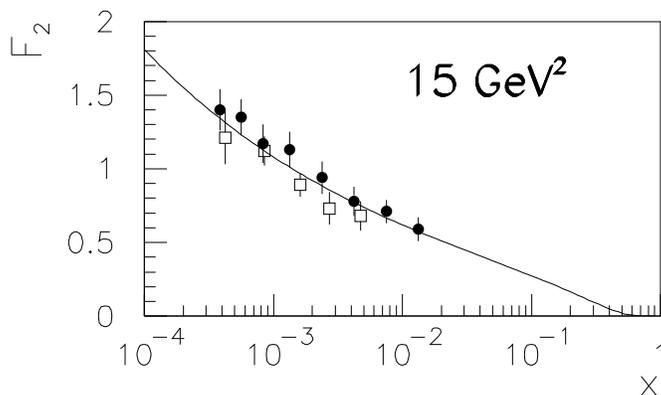
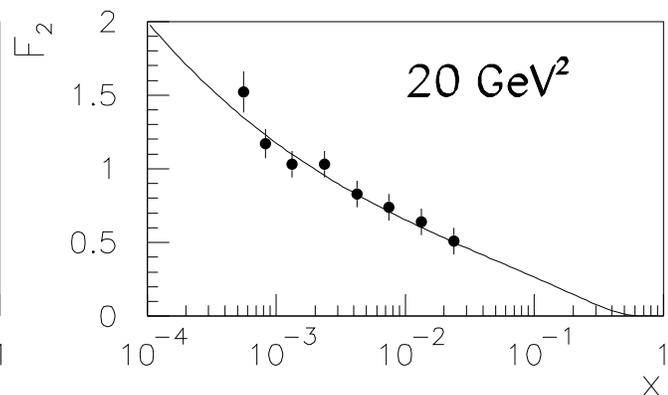
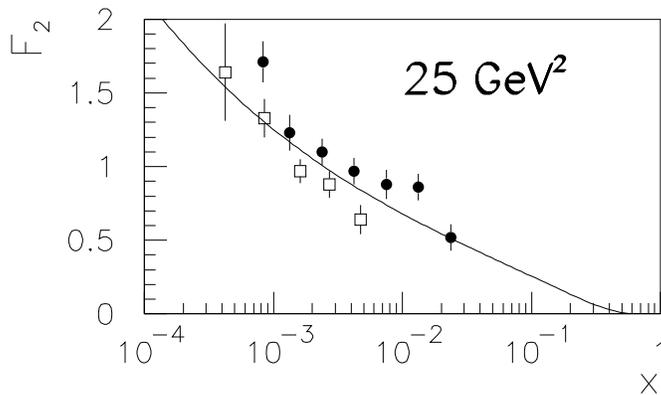
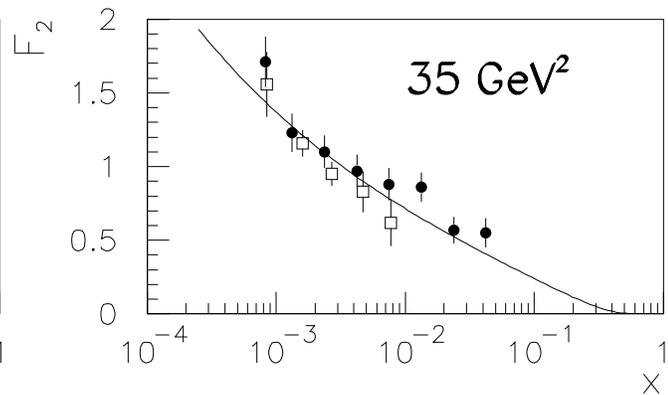
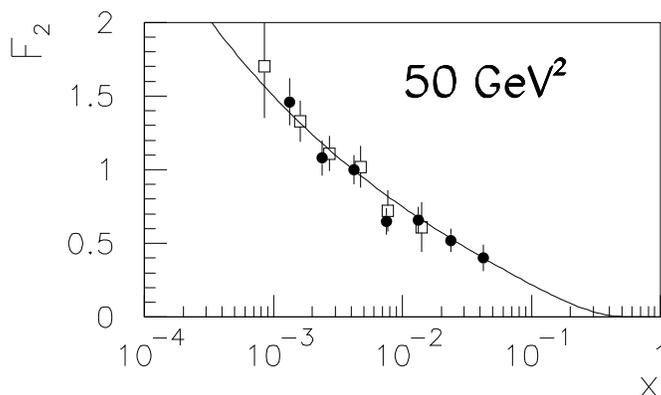
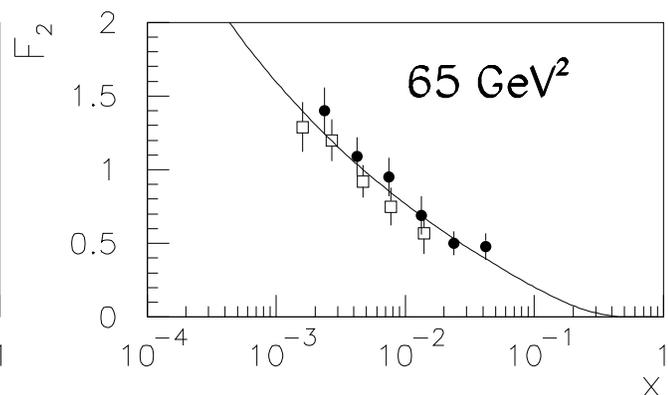
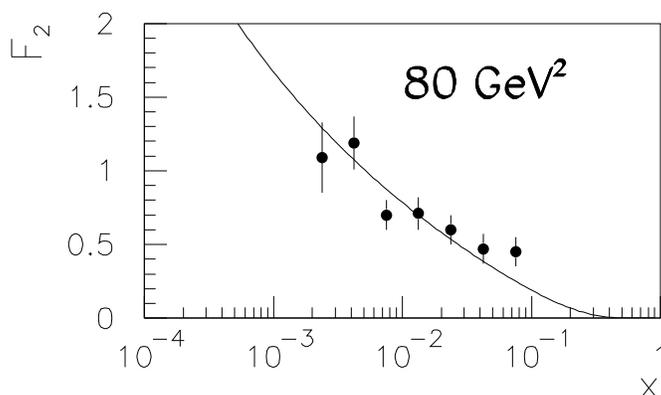
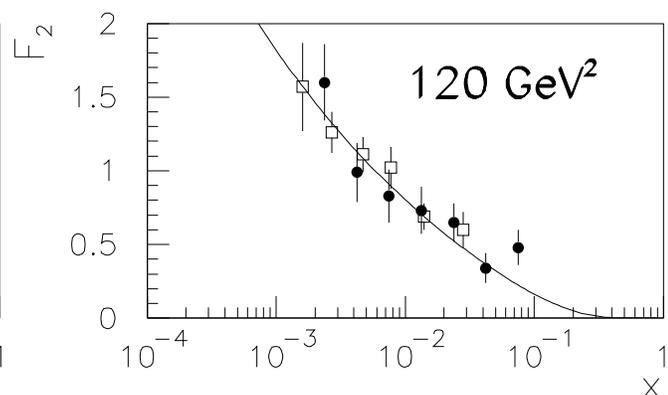
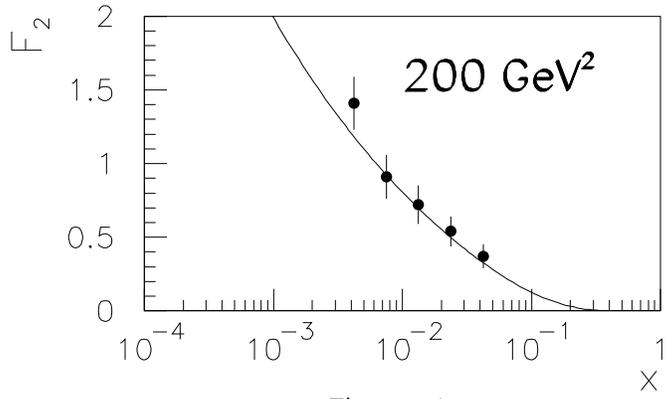
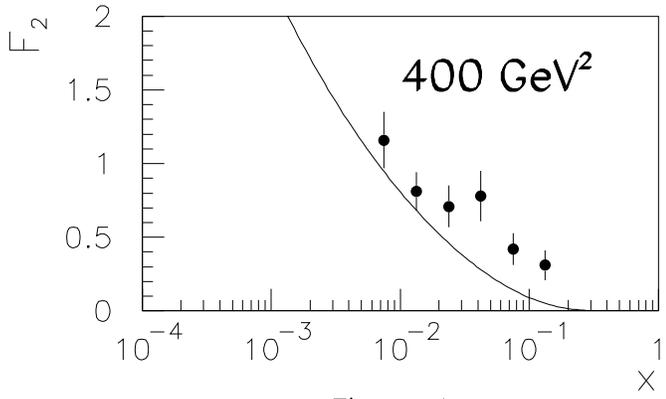
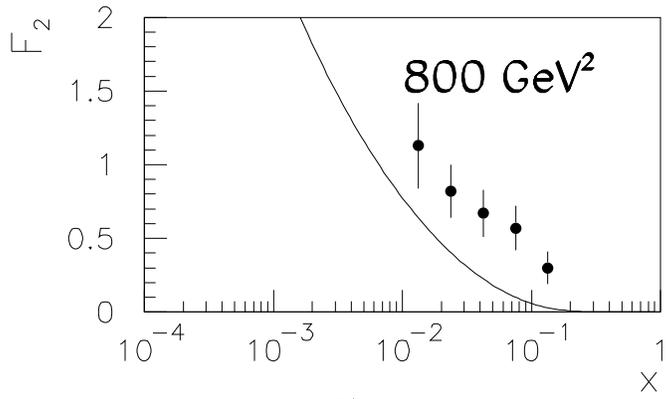
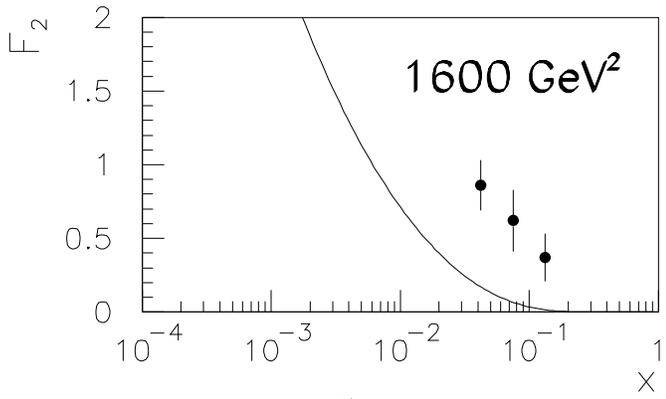
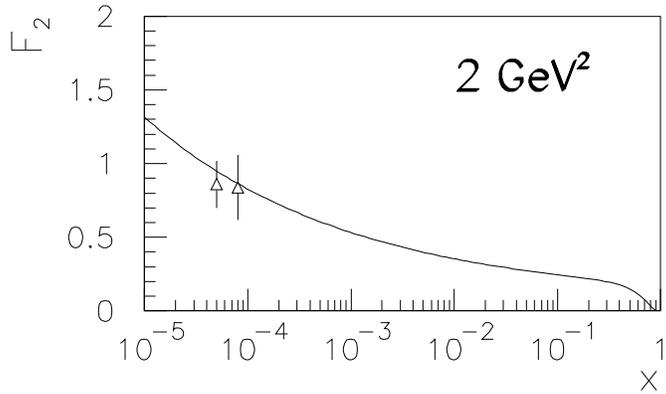
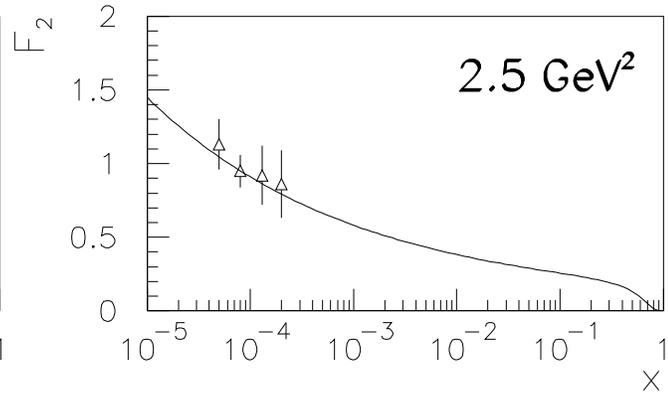
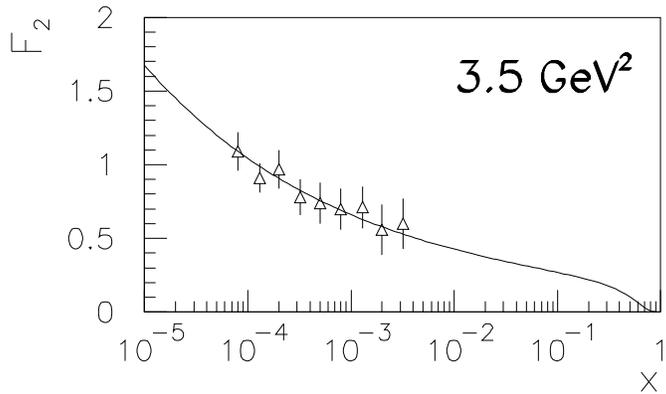
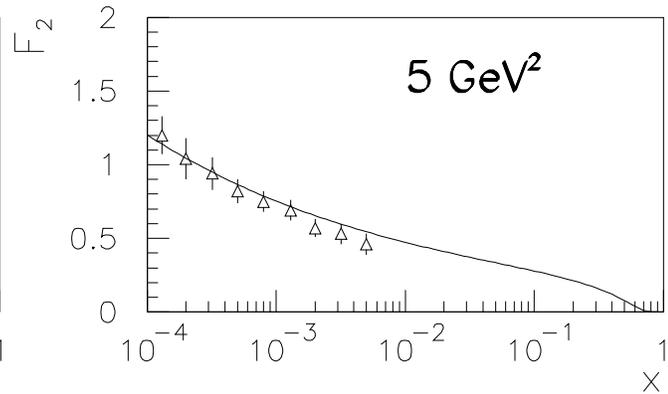
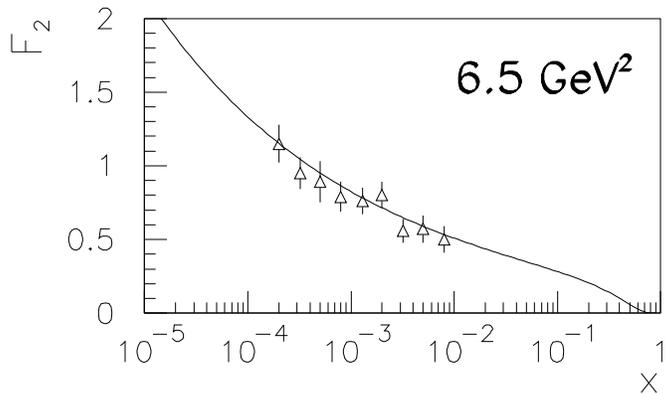
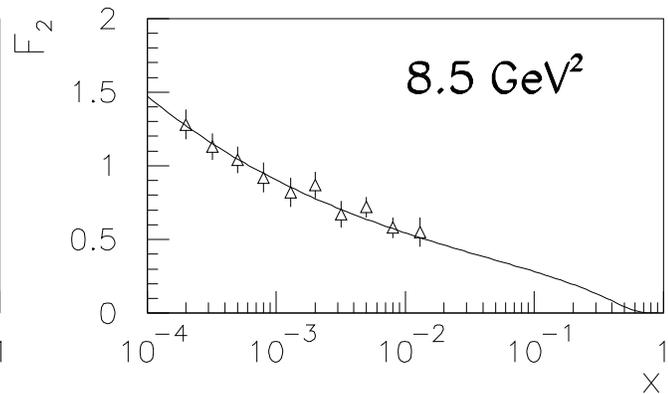
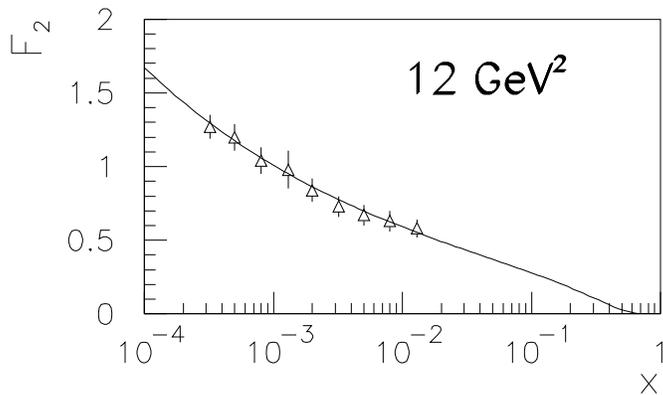
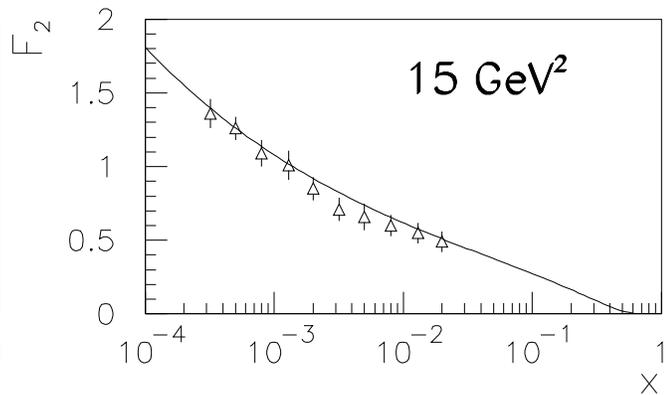
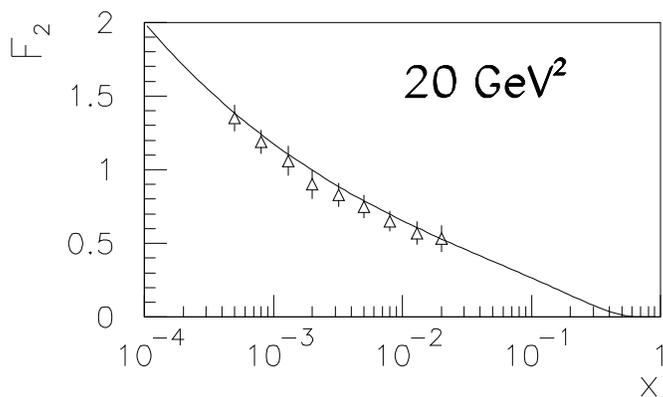
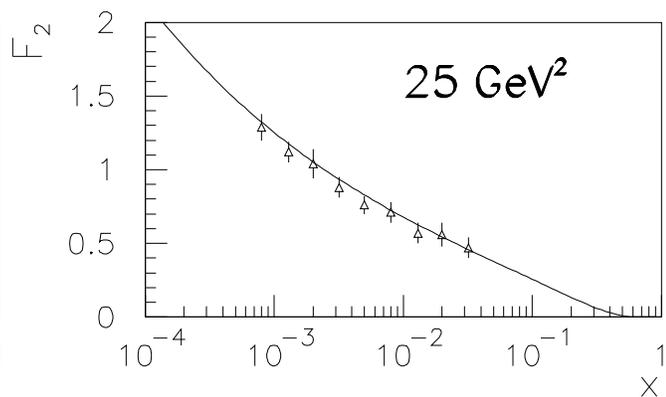
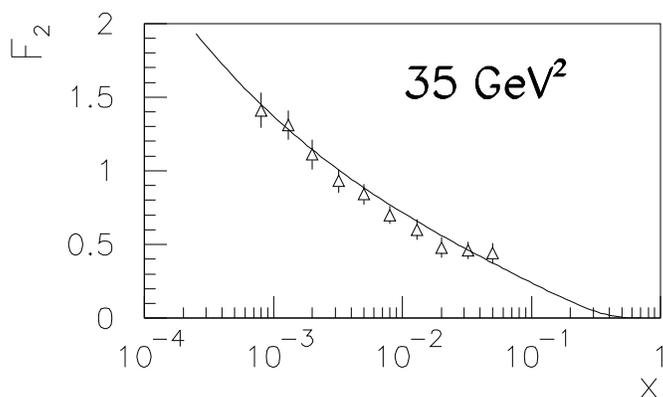
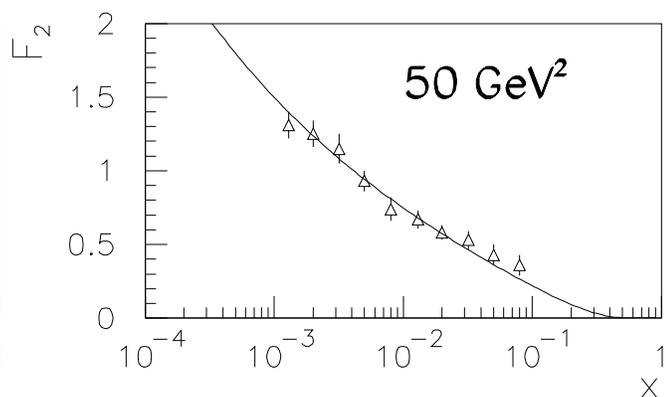
\end{document}